\documentclass[twocolumn,prl,amsmath,amssymb,showpacs,superscriptaddress,floatfix]{revtex4}
\usepackage{graphicx}
\usepackage{bm}

\usepackage{epsf}
\begin{document}
\title{Anomalous Melting Scenario of the Two-Dimensional Core-Softened System}
\author{D.E. Dudalov}
\affiliation{ Institute for High Pressure Physics RAS, 142190
Kaluzhskoe shosse, 14, Troitsk, Moscow, Russia}

\author{Yu.D. Fomin}
\affiliation{ Institute for High Pressure Physics RAS, 142190
Kaluzhskoe shosse, 14, Troitsk, Moscow, Russia}
\affiliation{Moscow Institute of Physics and Technology, 141700
Moscow, Russia}

\author{E.N. Tsiok}
\affiliation{ Institute for High Pressure Physics RAS, 142190
Kaluzhskoe shosse, 14, Troitsk, Moscow, Russia}

\author{V.N. Ryzhov}
\affiliation{ Institute for High Pressure Physics RAS, 142190
Kaluzhskoe shosse, 14, Troitsk, Moscow, Russia}
\affiliation{Moscow Institute of Physics and Technology, 141700
Moscow, Russia}

\date{\today}

\begin{abstract}
We consider the phase behavior of two-dimensional ($2D$)system of
particles with an isotropic core-softened potential introduced in
our previous publications. As one can expect from the qualitative
consideration for the three dimensional case, the system
demonstrates a reentrant-melting transition at low densities along
with waterlike anomalies in the fluid phase near the melting
maximum. In contrast with the three dimensional case, in two
dimensions melting is a continuous two-stage transition in the low
density part of the phase diagram with an intermediate hexatic
phase corresponding to the
Kosterlitz-Thouless-Halperin-Nelson-Young (KTHNY) scenario. At the
same time, at high densities the system melts through one
first-order transition. We also show, that the order of the region
of anomalous diffusion and the regions of density and structural
anomalies are inverted in comparison with the $3D$ case and have
silicalike sequence.
\end{abstract}
\maketitle

In recent years, a growing attention has been paid to
investigation of melting/freezing phenomena of confined fluids in
relation with the different fields of modern technology such as
fabrication of nanomaterials, nanotribology, adhesion, and
nanotechnology \cite{rev1,rice}. The fundamental question is how
the properties of a system change as the dimensionality changes
from three dimensions ($3D$) to two dimensions ($2D$). The most
interesting topics concern the existence of the specific $2D$
phase, hexatic phase, that interpolates between the fluid and
ordered solid phases, and the dependence of the nature of $2D$
phase transition on the character of the interparticle
interaction. In $3D$, systems melt through the first-order
transition due to the third-order term in the Landau expansion.
However, in $2D$ the singular fluctuations of the order parameter
(dislocations and disclinations) may cause  the qualitative
differences between $2D$ and $3D$ behavior of matter
\cite{rto1,rto2,RT1,RT2}.

Despite the long history of investigations, the melting transition
of most materials in $2D$ is not well understood, because theories
explaining the transition on a microscopic scale are not
available. Furthermore, the mechanism of melting depends on the
details of the interactions between the particles forming the
crystal lattice. In their pioneering works, Halperin, Nelson, and
Young \cite{halpnel79}, using the Kosterlitz-Thouless ideas
\cite{kosthoul73}, proposed the scenario of two-dimensional
melting which is fundamentally different from the melting scenario
of conventional three-dimensional systems. It has been shown that
the transition between a crystal and an isotropic liquid can occur
by means of two continuous transitions which correspond to
dissociation of bound dislocation and disclination pairs,
respectively. The low-temperature solid phase is characterized by
quasi-long-range translational order and long-range
bond-orientational order. Dislocations unbinding at some
temperature $T_m$ leads to a phase with short-range translational
order, but with quasi-long-range bond-orientational order. This
intermediate phase is called a hexatic phase. Paired disclinations
in the hexatic phase ultimately unbind themselves, driving a
second transition at a higher temperature $T_i$ into an isotropic
liquid.

This theory has strong support from experiments with electrons on
helium \cite{gram} and computer simulations of the 2D electron
systems \cite{strandburg88}. An experimental confirmation for the
KTHNY theory for crystal melting in $2D$ has been found in the
colloidal model system with repulsive magnetic dipole-dipole
interaction \cite{keim1,zanh,keim2}. However, a conventional
first-order transition between a two-dimensional solid and an
isotropic liquid is also a possibility (see, for example,
\cite{chui83,klein2,ryzhovTMP,ryzhovJETP}).

It should be noted that the KTHNY theory is phenomenological and
seems universal. It is not clear from this theory whether the
melting scenario depends on the shape of an intermolecular
potential. Actually, the natural way to analyze this dependence is
to use computer simulations. However, simulations are not reliable
enough in the case of two-dimensional melting: it is interesting
to note that the similar simulation methods have led to
contradictory conclusions even when applied to the same systems
\cite{strandburg92,binderPRB,mak,jaster2,andersen,hfo1,binder,DF,LL,prest1,prest2}.
The problems are understandable since correlation times and
lengths (translational and orientational) can be extremely long
near the phase transition. A lot of efforts were made on
computational studies of two-dimensional melting of hard-core
potential systems including hard disks or Lennard-Jones potentials
\cite{strandburg92,binderPRB,mak,jaster2,andersen,hfo1,binder}.
Simulation results on these systems tend to favor a first-order
transition scenario for melting, although some conflicting results
also exist. In spite of all these efforts, a satisfactory answer
has not been obtained yet for one of the most important questions
in two-dimensional melting, which is as follows: what condition
determines the existence of a hexatic phase and the nature of the
melting transition? It seems natural to relate this behavior with
the range and the softness of the potential
\cite{LL,prest1,prest2}.

It is well known that some liquids demonstrate anomalous behavior
in some regions of thermodynamic parameters. The most common and
well known example is water. The water phase diagrams have regions
where a thermal expansion coefficient is negative (density
anomaly), self-diffusivity increases upon compression (diffusion
anomaly), and the structural order of the system decreases with
increasing pressure (structural anomaly). Later on it was
discovered that many other substances also demonstrate similar
behavior. Some typical examples are silica, silicon, phosphorus,
and many others. It is reasonable to relate this kind of behavior
to the orientational anisotropy of the potentials, however, a
number of studies demonstrate waterlike anomalies in fluids that
interact through spherically symmetric core-softened potentials
with two length scales. A lot of different core-softened
potentials were introduced (see, for example, reviews
\cite{buld2009,fr1}). However, it should be noted that in general
the existence of two length scales is not enough to mark the
occurrence of the anomalies. For example, for the models studied
in Ref. \cite{prest2,prest3} it was shown that the existence of
two distinct repulsive length scales is not a necessary condition
for the occurrence of anomalous phase behavior.

In this work, we present a simulation study of two-dimensional
melting transition and anomalous behavior in the purely repulsive
core-softened system introduced in our previous publications
\cite{wejcp,wepre,we_inv,we2013,RCR,we2013-2}. The general form of
the potential is written as
\begin{equation}
U(r)=\varepsilon\left(\frac{\sigma}{r}\right)^{14}+\frac{1}{2}\varepsilon\left(1-
\tanh(k_1\{r-\sigma_1\})\right). \label{3}
\end{equation}
Here $k_1=10.0$, and $\sigma_1=1.35$. In the remainder of this
paper we use the dimensionless quantities: $\tilde{{\bf r}}\equiv
{\bf r}/\sigma$, $\tilde{P}\equiv P \sigma^{2}/\varepsilon ,$
$\tilde{V}\equiv V/N \sigma^{2}\equiv 1/\tilde{\rho}, \tilde{T}
\equiv k_BT/\varepsilon $. Since we will use only these reduced
variables, the tildes will be omitted.

In $3D$, particles interacting through a purely repulsive
potential given by Eq.~(\ref{3}) exhibit reentrant melting, a
maximum melting temperature, superfragile glass behavior, and
anomalies similar to the ones found in water and silica
\cite{wejcp,wepre,we_inv,we2013,RCR,we2013-2}.

As it was discussed before \cite{RT1,RT2}, there are two
characteristic temperatures for the melting transition in $2D$:
the dislocation unbinding temperature $T_m$ and the first-order
transition temperature $T_{MF}$. The modulus of the order
parameter vanishes at the temperature $T_{MF}$ which can be
obtained from the double-tangent construction for the free
energies of liquid and solid phases. There are two possibilities
\cite{RT1,RT2}: 1: $T_m<T_{MF}$. In this case the system melts via
two continuous transitions of the Kosterlitz-Thouless type with
the unbinding of dislocation pairs. 2: $T_m>T_{MF}$. The system
melts via a first-order transition because of the existence of
third-order terms in the Landau expansion as in the ordinary
three-dimensional case \cite{RT1,RT2}. The phase diagram
corresponding to $T_{MF}$, gives the limit of the thermodynamic
stability of the solid phase. In order to conclude whether the
melting occurs through the KTHNY scenario, the additional analysis
is necessary.

We simulate the system in $NVT$ and $NVE$ ensembles using the
molecular dynamics  (LAMMPS package \cite{lammps}). The number of
particles in the simulation varied between $3200$ and $102400$. In
order to find the transition points we carry out the free energy
calculations for different phases and construct a common tangent
to them. For the purely repulsive potentials we computed the free
energy of the liquid by integrating the equation of state along an
isotherm \cite{book_fs}:
$\frac{F(\rho)-F_{id}(\rho)}{Nk_BT}=\frac{1}{k_BT}\int_{0}^{\rho}\frac{P(\rho')-\rho'
k_BT}{\rho'^2}d\rho'$. Free energies of different crystal phases
were determined by the method of coupling to the Einstein crystal
\cite{book_fs}. The phase diagram calculated in this way
corresponds to the first-order transitions scenario.

\begin{figure}
\includegraphics[width=4cm]{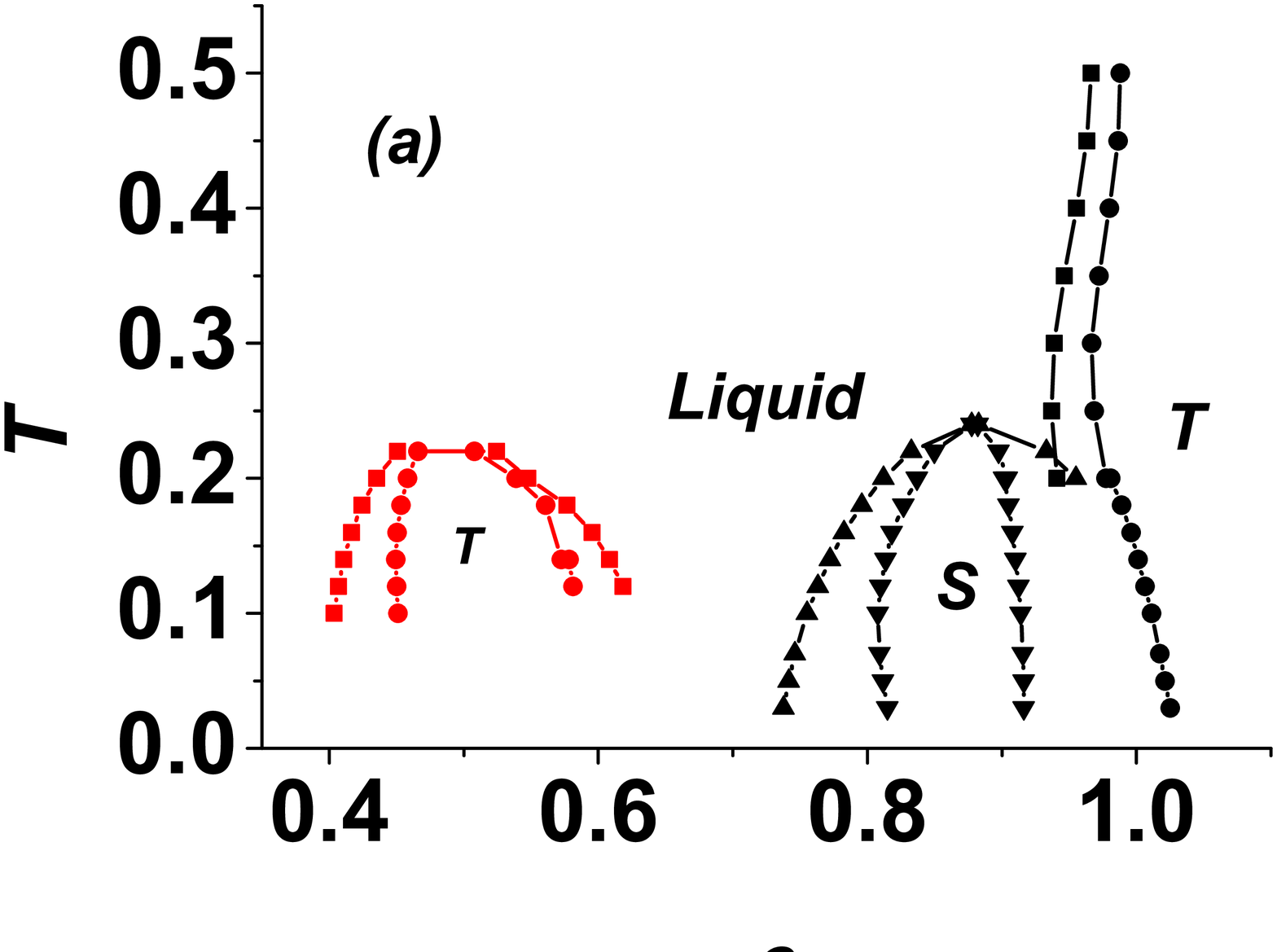}
\includegraphics[width=4cm]{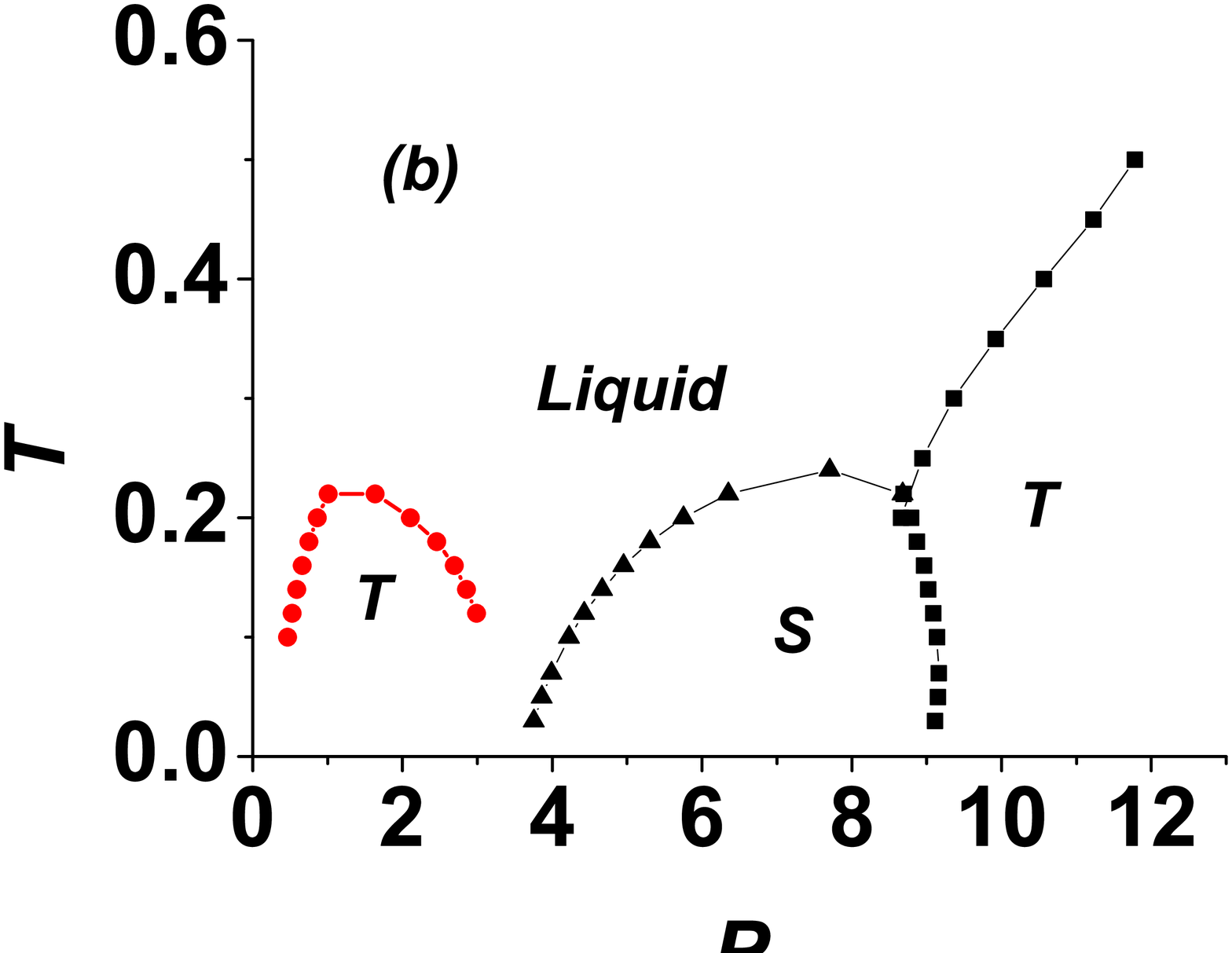}

\caption{\label{fig:fig2} (a) Phase diagram of the system with the
potential (\ref{3}) in $\rho-T$ plane, where the Triangular (T)
and Square (S) phases are shown. (b) Phase diagram of the same
system in the $P-T$ plane.}
\end{figure}

We plot in Fig.~\ref{fig:fig2} the phase diagram of the system in
$\rho-T$ and $P-T$ coordinates. There is a clear maximum in the
melting curve at low densities. The phase diagram consists of two
isostructural triangular crystal domains (T) corresponding to
close packing of the small and large disks separated by a
structural phase transition and square lattice (S). Similar phase
diagram in $3D$ was discussed in details in our previous
publications \cite{wejcp,wepre}. It is important to note that
there is a region of the phase diagram where we have not found any
stable crystal phase. The results of $3D$ simulations
\cite{wejcp,RCR} suggest that a glass transition can occur in this
region.

\begin{figure}
\includegraphics[width=4cm]{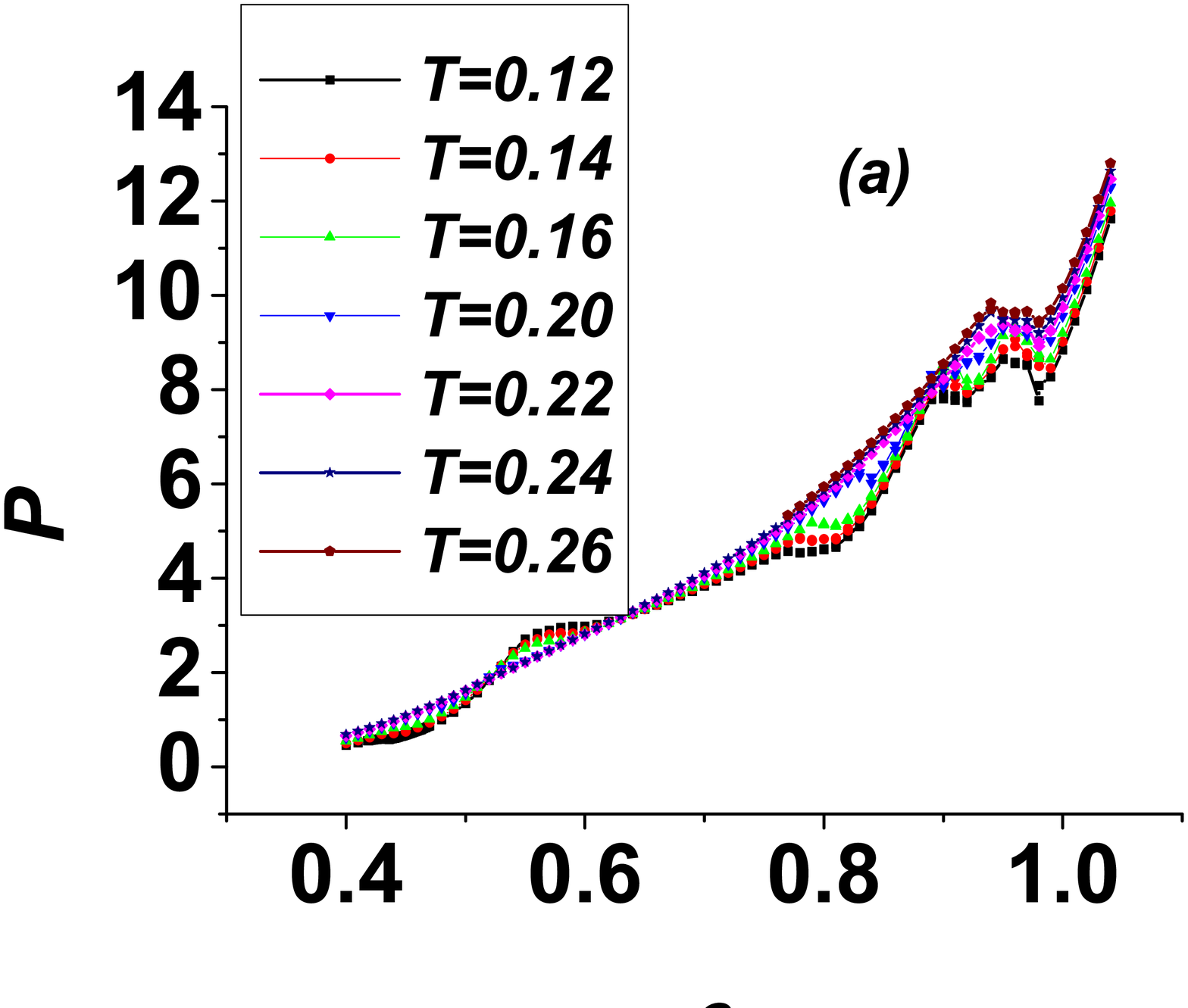}
\includegraphics[width=4cm]{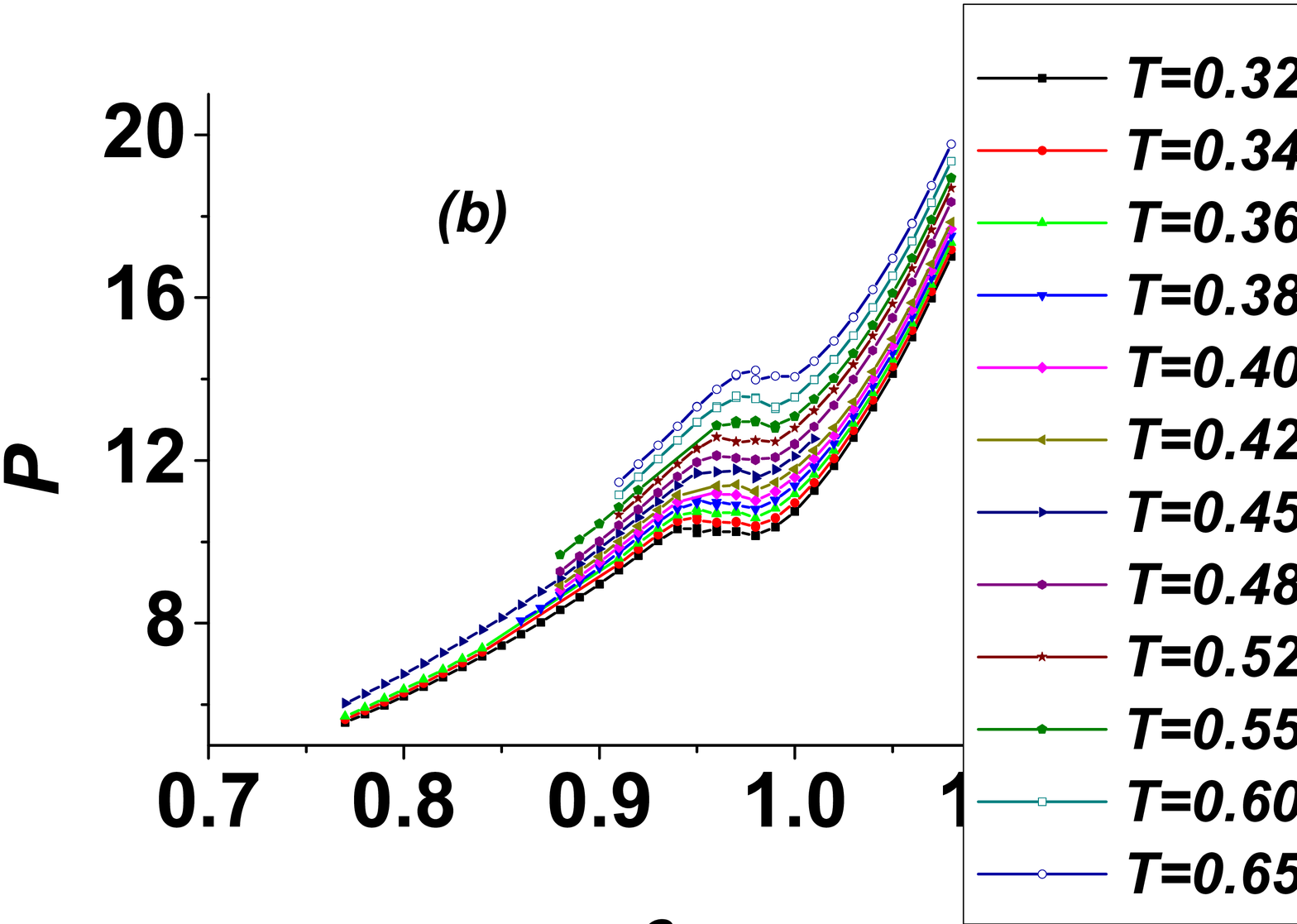}

\caption{\label{fig:fig3} (Color online) The low-temperature (a)
and high-temperature (b) sets of isotherms.}
\end{figure}

To distinguish the first-order melting scenario from the
continuous one, we used the criteria described in the Ref.
\cite{binder}. In Fig.~\ref{fig:fig3} we present the
low-temperature (Fig.~\ref{fig:fig3}(a)) and high-temperature
(Fig.~\ref{fig:fig3}(b)) sets of isotherms. One can see that at
low temperatures there are four regions on the isotherms
corresponding to the phase transitions (see
Fig.~\ref{fig:fig3}(a)), the low density ones being smooth as in
the case of liquid-hexatic-solid transition \cite{binder} and the
high densities  part containing the Van der Waals loops
characteristic of the first order phase transition. At high
temperatures (see Fig.~\ref{fig:fig3}(b)) there is only one
liquid-triangular lattice first-order transition. From
Fig.~\ref{fig:fig3} one can guess that the melting of the
low-density and high-density parts of the phase diagram occurs
with different scenarios: at low densities the KTHNY scenario is
probable, while the high density phase melts through the
first-order phase transition. As we are going to show in the
following, the intermediate region between the solid and the
(normal) fluid can be qualified as hexatic.

To confirm this guess, let us define the translational order
parameter $\psi_T$ (TOP), the orientational order parameter
$\Psi_n$ (OOP), and the bond-orientational correlation function
$G_n(r)$ (OCF) in the conventional way
\cite{prest2,halpnel79,strandburg88,binderPRB,binder,prest1}.

TOP an be used in the form
\begin{equation}
\psi_T=\frac{1}{N}\left<\left|\sum_i e^{i{\bf G
r}_i}\right|\right>, \label{psit}
\end{equation}
where ${\bf r}_i$ is the position vector of particle $i$ and {\bf
G} is the first reciprocal-lattice vector. It may be easily seen
that $\psi_T$ is nonzero if a solid the orientation corresponding
to the length and direction of {\bf G}. In the simulation, melting
of the crystal phase into hexatic phase or isotropic liquid is
determined by the sharp decrease of $\psi_T$ on heating.

The orientational order and the hexatic phase may be studied with
the help of the local order parameter, which can be used for
measuring the $n$-fold orientational ordering:
\begin{equation}
\Psi_n({\bf r_i})=\frac{1}{n(i)}\sum_{j=1}^{n(i)} e^{i
n\theta_{ij}}\label{psi6l},
\end{equation}
where $\theta_{ij}$ is the angle of the bond joining the particles
$i$ and $j$ with respect to a reference axis and the sum over $j$
is over all $n(i)$ nearest-neighbors of $j$. The Voronoi
construction is used to obtain $n(i)$.  An average over all
particles gives the global OOP:
\begin{equation}
\psi_n=\frac{1}{N}\left<\left|\sum_i \Psi_n({\bf
r}_i)\right|\right>.\label{psi6}
\end{equation}
It should be noted that $n=6$ corresponds to the triangular solid
and $n=4$ - to square solid. In a perfect triangular solid
$n(i)=6$, $\theta_{ij}=\pi/3$ and $\psi_6=1$.

The bond-orientational correlation function $G_n(r)$ (OCF) is
given by the equation:
\begin{equation}
G_n(r)=\left<\Psi_n({\bf r})\Psi_n^*({\bf 0})\right>, \label{g6}
\end{equation}
where $\Psi_n({\bf r})$ is the local bond-orientational order
parameter (\ref{psi6l}).

In the isotropic fluid phase and in the hexatic phase,
$\psi_n\rightarrow 0$ as $L\rightarrow \infty$, where $L$ is the
linear size of the system, but the behaviors of $G_n(r)$ are
different in hexatic and isotropic phases. In the framework of the
KTHNY theory, an algebraic large-distance decay of the OCF is
predicted for the hexatic phase, in contrast with the exponential
asymptotic decay of angular correlations in a normal isotropic
fluid:
\begin{eqnarray}
G_n(r)&=&e^{-r/\xi}, r\rightarrow \infty, \rho<\rho_l,
\label{g61}\\
G_n(r)&=&r^{-\eta(T)}, r\rightarrow \infty, \rho_l<\rho<\rho_s
\label{g62}.
\end{eqnarray}
Here $\xi$ is the correlation length of the bond orientational
order, which diverges as $\rho_l$ is approached. Another
prediction of the theory is $\eta=1/4$ at the hexatic-to-normal
isotropic fluid transition point \cite{halpnel79}.

The corresponding susceptibility \cite{binderPRB,binder}
\begin{equation}
\chi_n=\frac{1}{N}\left<\left|\sum_i \Psi_n({\bf
r}_i)\right|^2\right>-N\psi_n^2, \label{chi}
\end{equation}
shows a peak. Location of the peak estimates the transition point.

\begin{figure}
\includegraphics[width=4cm]{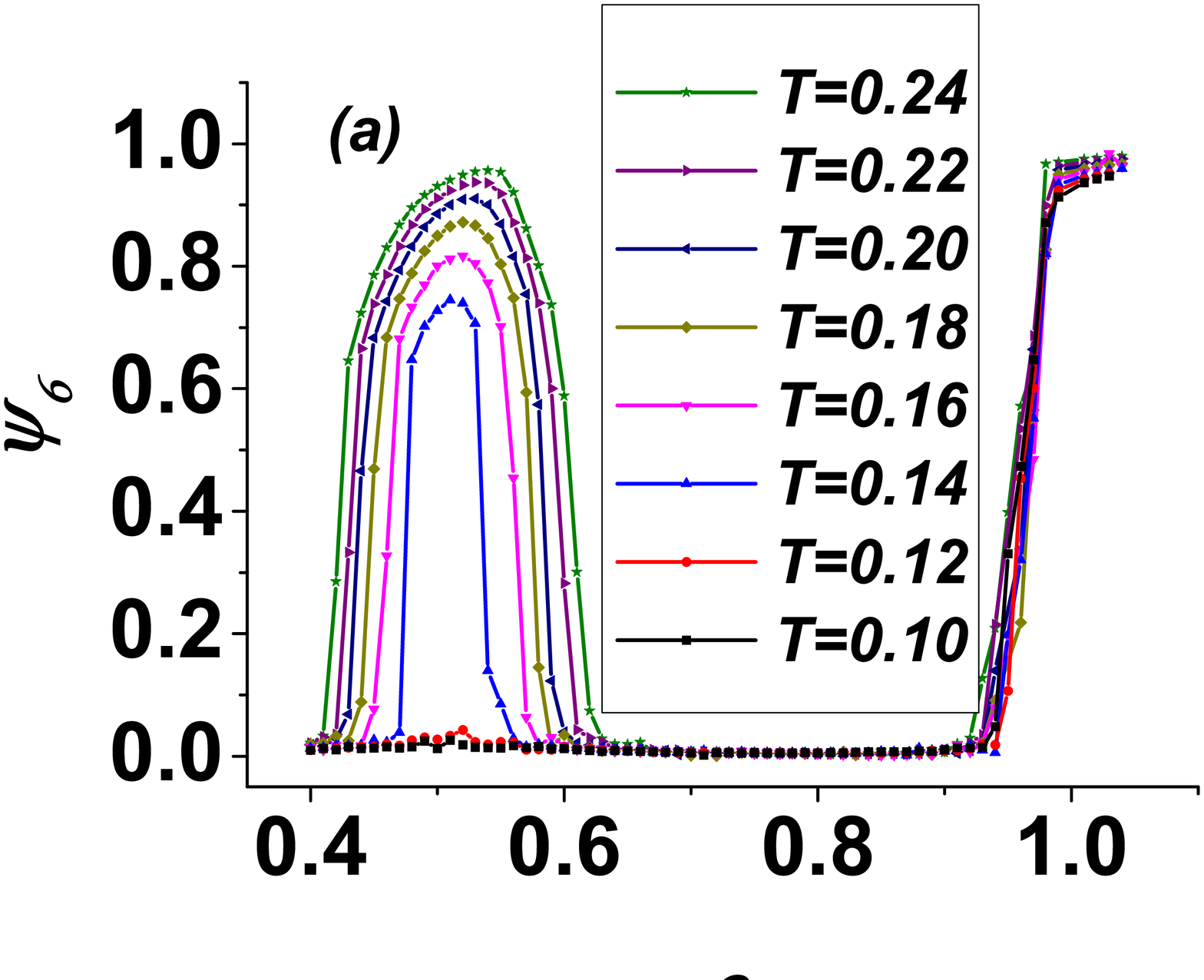}%
\includegraphics[width=4cm]{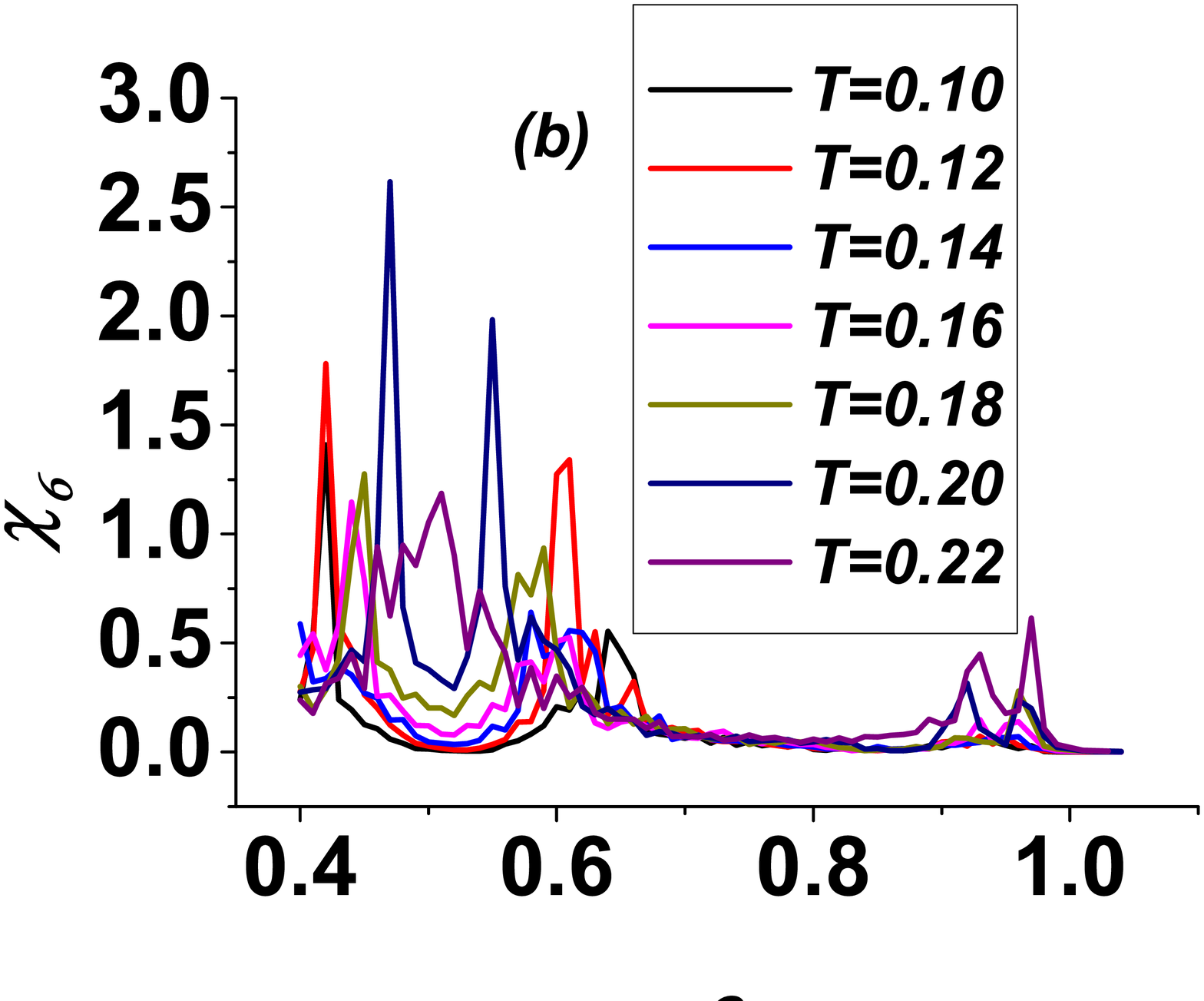}%

\caption{\label{fig:fig4} (Color online) (a) Orientational order
parameter as a function of density for different temperatures; (b)
The corresponding susceptibility $\chi_6$ as a function of density
for different temperatures.}
\end{figure}

In Fig.~\ref{fig:fig4}(a), we represent the orientational order
parameter (OOP) as a function of density for a set of
temperatures. We see, that at the low density part of the phase
diagram OOP behaves smoothly while at high densities one can see
the abrupt change of OOP. This kind of behavior suggests again
that the melting at low densities is continuous in accordance with
the KTHNY scenario, and at high densities melting transition is of
the first order. In Fig.~\ref{fig:fig4}(b), the corresponding
susceptibility is shown as a function of density for several
temperatures. One can see, that at low densities, $\chi_6$
demonstrates the sharp peaks characteristic for the continuous
transition, while at high densities the peaks are much smaller, as
in the case of the first-order phase transition.

\begin{figure}
\includegraphics[width=4cm]{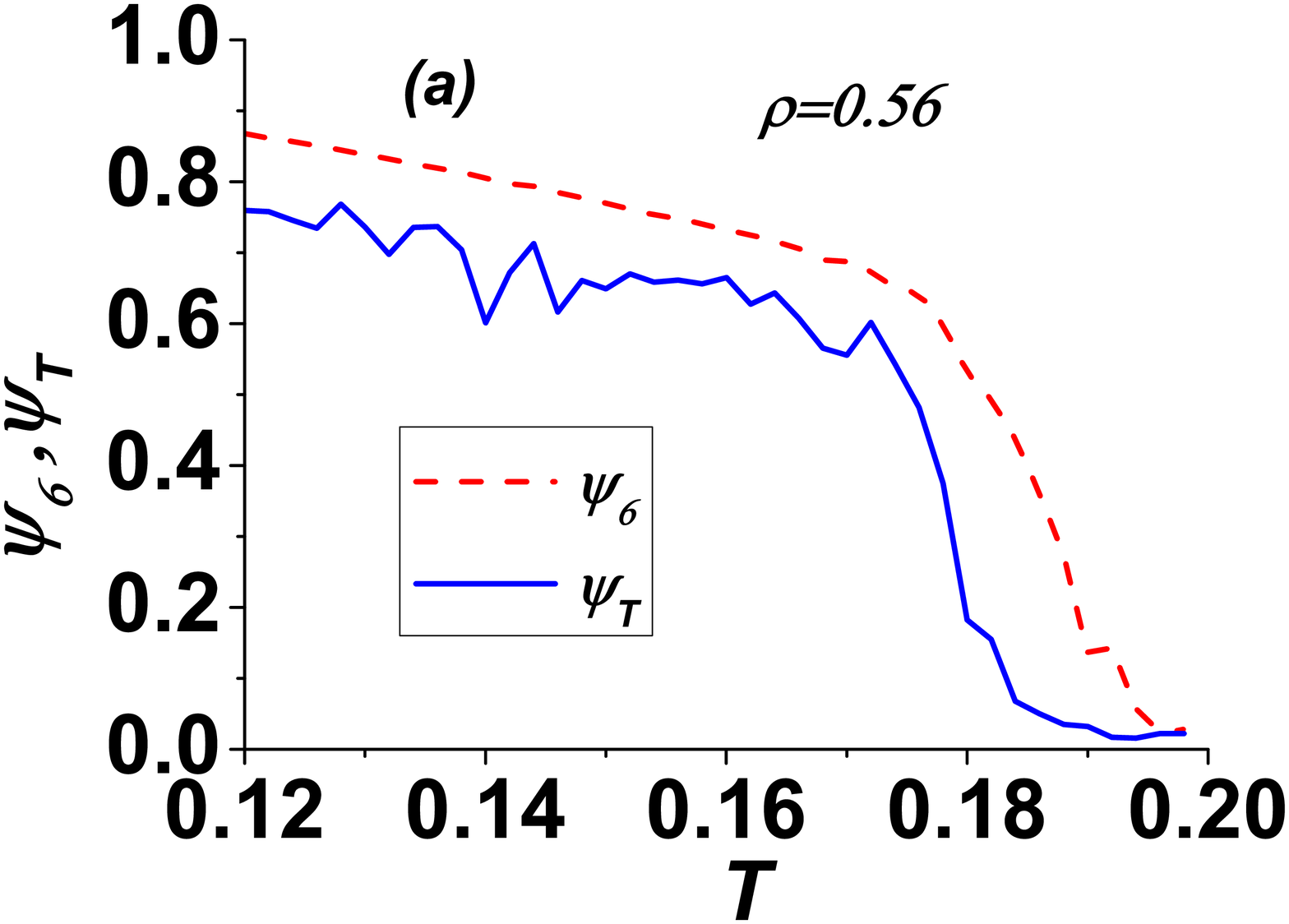}%
\includegraphics[width=4cm]{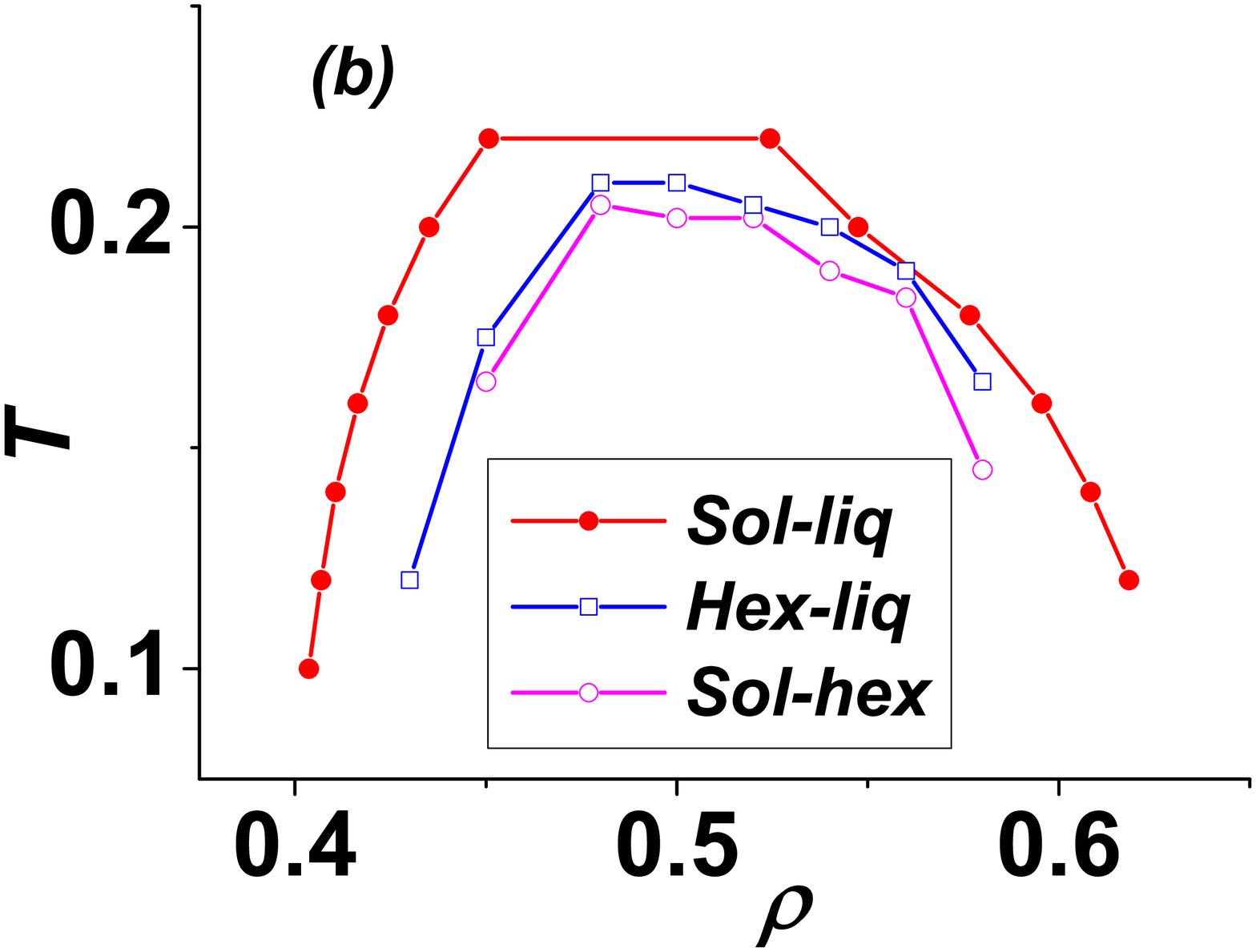}%

\caption{\label{fig:fig5} (Color online) (a). OPs $\psi_T$ and
$\psi_6$ as functions of temperature for $\rho=0.56$. It is
clearly the narrow hexatic phase; (b) The low-density part of the
phase diagram (Fig.~\ref{fig:fig2}(a)) along with the lines of
solid-hexatic and hexatic-liquid transitions. }
\end{figure}

In Fig.~\ref{fig:fig5}(a), we plot the two OPs for $\rho=0.56$ as
a function of temperature (an analogous behavior was observed for
all the other densities). We see that $\psi_T$ vanishes at a
slightly smaller temperature than $\psi_6$, which implies that the
hexatic phase is confined to an  narrow T interval.  In
Fig.~\ref{fig:fig5}(b), the phase transition line of the
solid-hexatic and hexatic-liquid transitions are show in
comparison with the solid-liquid transition line (see
Fig.~\ref{fig:fig2}). One can see that the transitions are mainly
inside the solid region, obtained in the framework of the
free-energy calculations. This fact also supports the idea that
the melting in this region occurs through two continuous
transitions. It is necessary to note, that in the case of the
conventional first-order phase transition, the density change at
the melting line maximum is equal to zero. We see, that the
hexatic phase becomes narrower in the vicinity of the maximum,
however, our calculations can not answer whether the width of
hexatic region tends to zero at the maximum point.

\begin{figure}
\includegraphics[width=8cm]{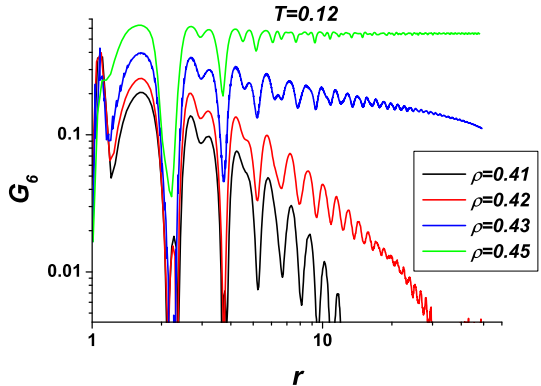}

\caption{\label{fig:fig7} (Color online) Log-log plots of the
orientational correlation function $G_6(r)$ at selected densities
across the hexatic region for $T=0.12$. Upon increasing $\rho$
from 0.41 to 0.45 there is a qualitative change in the
large-distance behavior of $G_6(r)$, from constant (solid) to
power-law decay (hexatic fluid), up to exponential decay (normal
fluid). Note that, consistently with the KTHNY theory, the decay
exponent $\eta$ is less than $1/4$ for $\rho>0.43$.}
\end{figure}

In order to get another evidence of the hexatic phase existence
can be obtained from the large-distance behavior of the OCF. It is
shon in Fig.~\ref{fig:fig7} for several densities for $T=0.12$.
One can see that the OCF decays algebraically in a $\rho$ region,
approximately corresponding to the hexatic phase region in
Fig.~\ref{fig:fig5}(b).

\begin{figure}
\includegraphics[width=4cm]{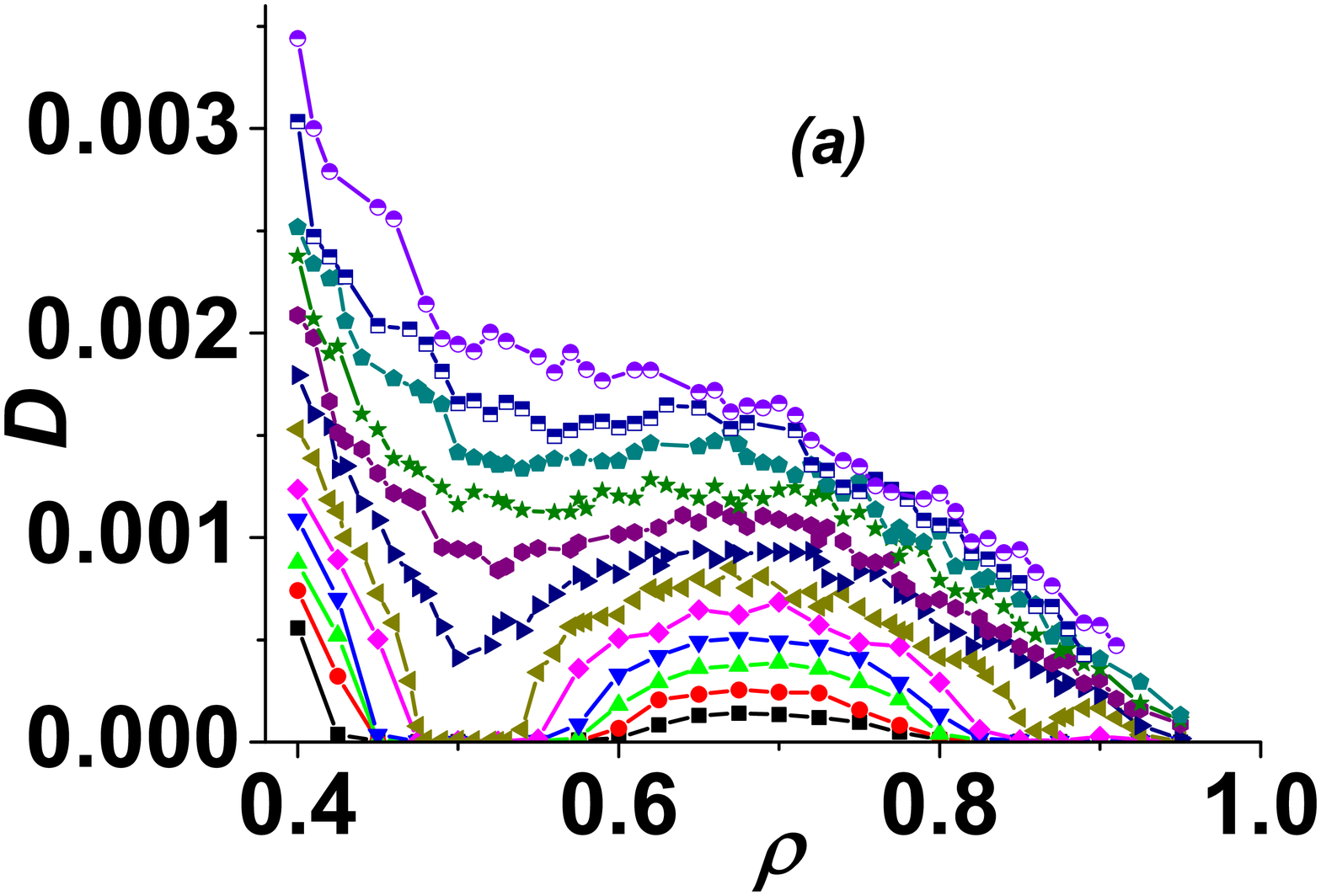}%
\includegraphics[width=4cm]{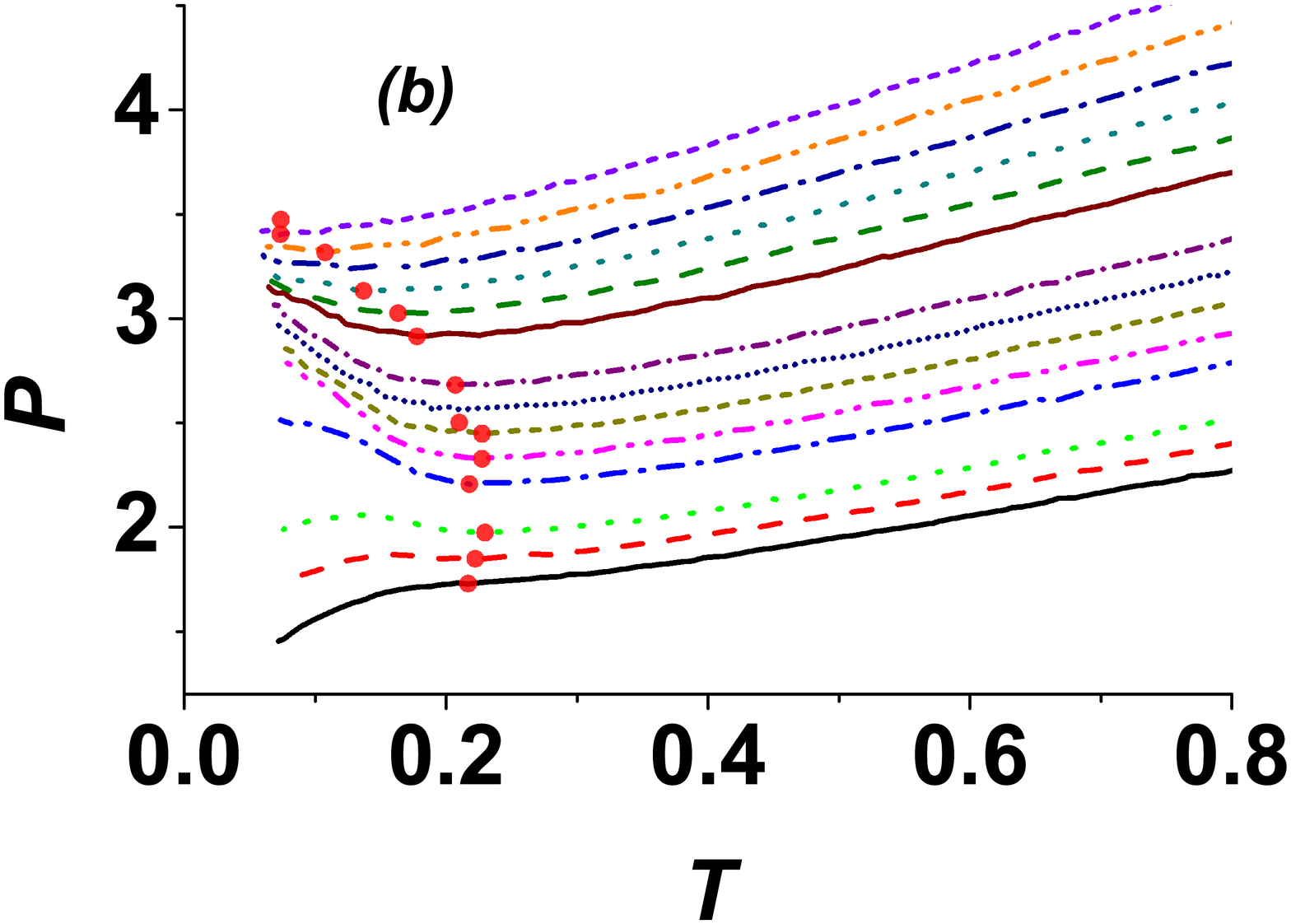}%

\includegraphics[width=4cm]{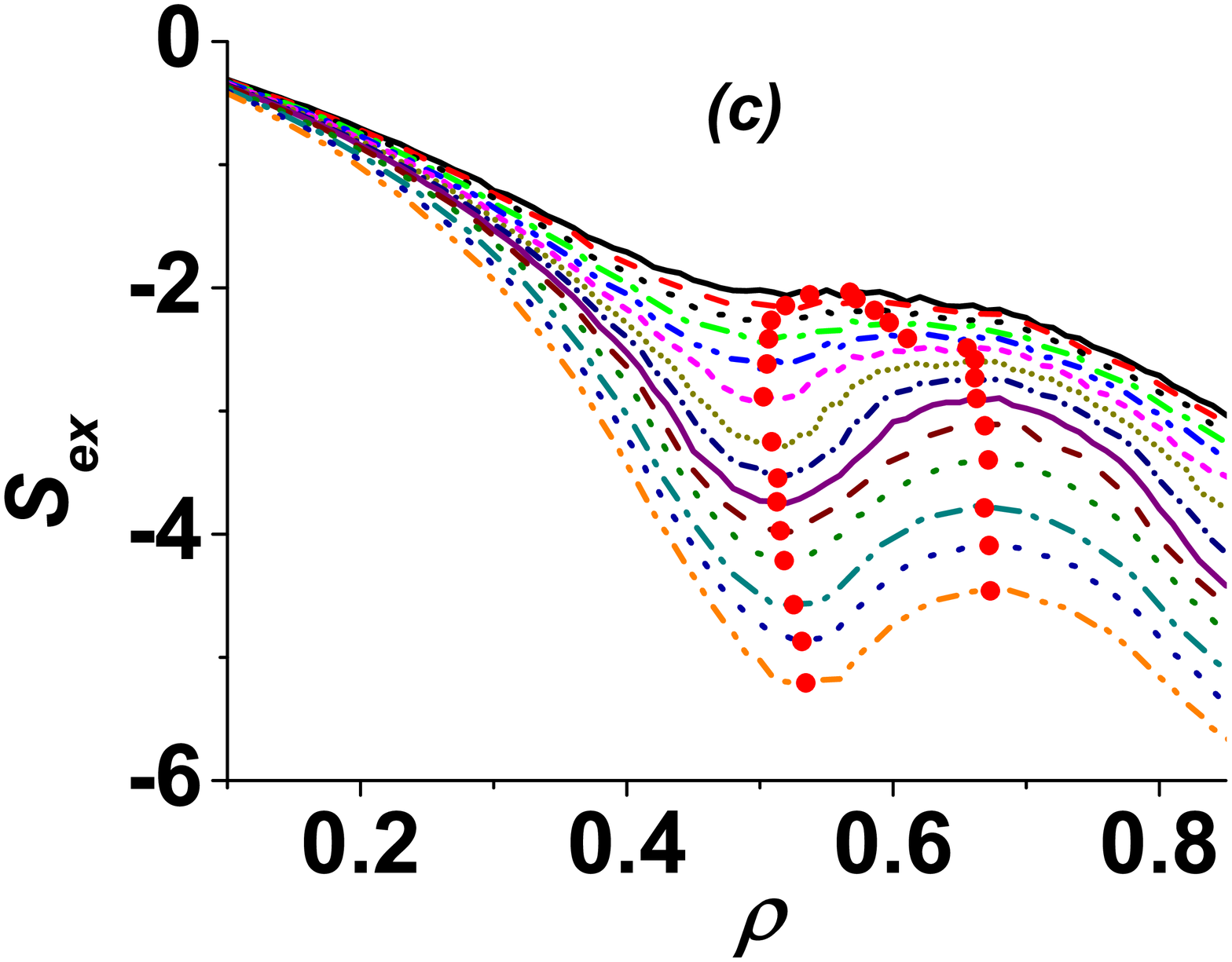}%
\includegraphics[width=4cm]{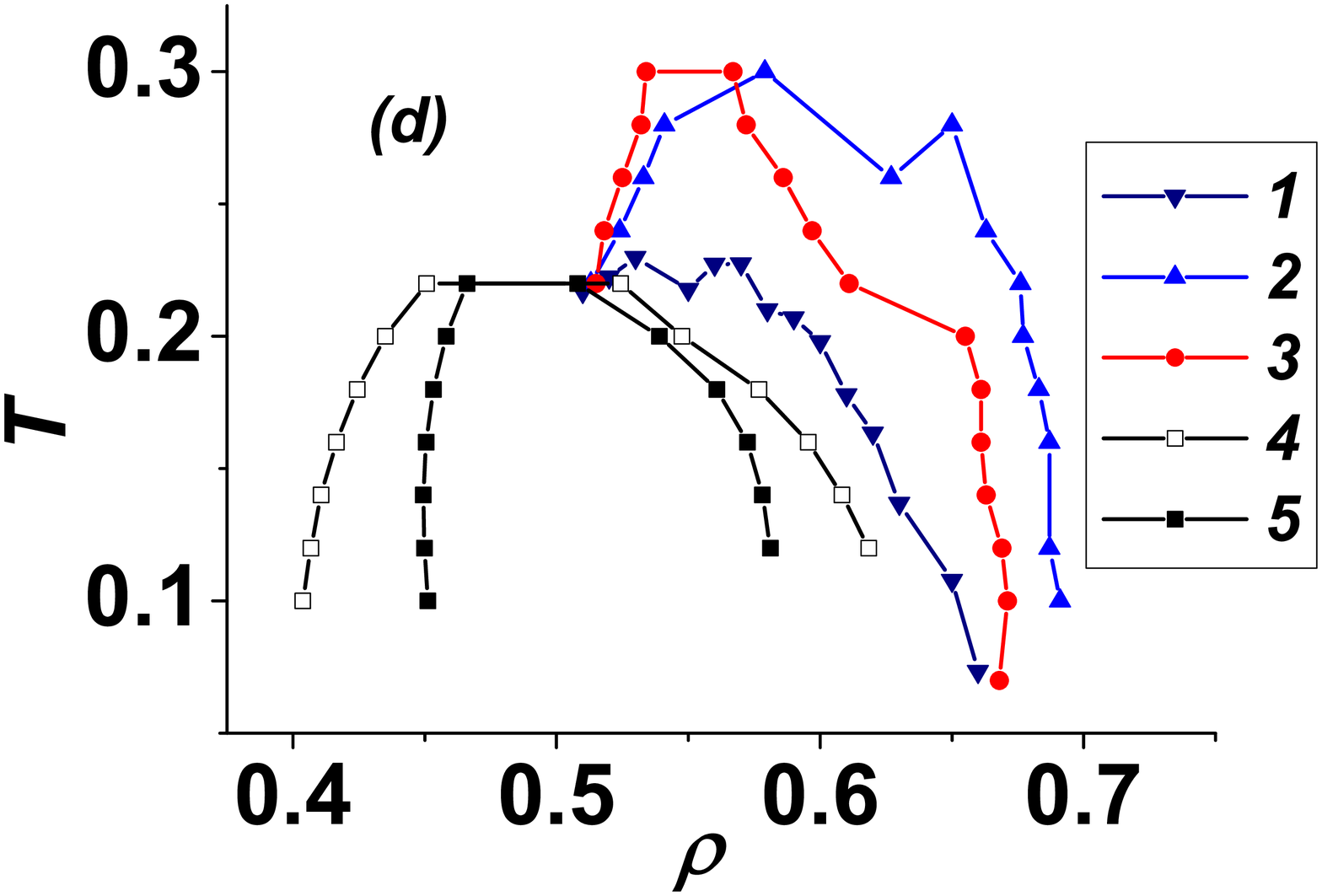}%

\caption{\label{fig:fig8} (Color online) (a) Diffusion coefficient
along a set of isotherms as a function of density. Results are for
temperatures $T=$ 0.1, 0.12, 0.14, 0.18, 0.20, 0.22, 0.24, 0.26,
0.28, 0.30, 0.32 from bottom up.  At low temperatures, there are
regions where the diffusion coefficient is increasing with
increasing density (diffusion anomaly); (b) pressure along a set
of isochors as a function of temperature. The lines correspond to
densities $\rho=$ 0.51, 0.52, 0.53, 0.55, 0.56, 0.57, 0.58, 0.59,
0.61, 0.62, 0.63, 0.64, 0.65, 0.66 from bottom to top. Minimum on
isochores corresponds to the density anomaly; (c) excess entropy
$S_{ex}$ along a set of isotherms as a function of density.
$S_{ex}=S-S_{id}$ is equal to the difference between the total $S$
and ideal gas $S_{id}$ entropies. Results are for temperatures
$T=$ 0.03, 0.05, 0.07, 0.10, 0.12, 0.14, 0.16, 0.18, 0.20, 0.22,
0.24, 0.26, 0.28, 0.30 from bottom up. At low temperatures, there
are regions where the excess entropy is increasing with increasing
density (structural anomaly); (d) the boundaries of anomaly
regions: 1). isobaric $\rho$ maxima (density anomaly); 2).
isothermal $D$ minima and maxima (left and right blue triangles);
3). isothermal $S_{ex}$ maxima and minima (left and right red
balls); 4). and 5). the borders of the low-density triangle phase.
The silicalike \cite{we2013,we2013-2,silica} order of anomalies
takes place: the diffusion anomaly region contains the
structurally anomalous region which, in turn, incorporates the
density anomaly region.}
\end{figure}

It should be noted, that the scaling analysis also supports the
melting scenario described above. The similar analysis was made
for the melting of the square lattice region of the phase diagram,
and it was shown that the square lattice melts through the
first-order phase transition.

It was shown that the mentioned above anomalous behavior also
exists in $2D$ \cite{scala,barbosa, buld2d,krott}. The
core-softened systems, described by the potential (\ref{3}),
demonstrate the anomalous behavior in three dimensions
\cite{wejcp,wepre,we_inv,we2013,RCR,we2013-2}. In $2D$, we found
the same anomalies (see Fig~\ref{fig:fig8}), however, the order of
the region of anomalous diffusion and the regions of density and
structural anomalies are inverted in comparison with the $3D$ case
and have silicalike sequence \cite{we2013,we2013-2,silica}. It
should be noted, that the similar sequence of anomalies was found
in Ref. \cite{prest2} for extremely soft potential, however, the
authors of Ref. \cite{prest2} did not compare the  $2D$ and $3D$
cases.

In conclusion, we show that at low densities the core softened
system defined by the potential Eq. (\ref{3}) demonstrates a
two-stage continuous reentrant melting via a hexatic phase. At the
same time, at high densities the system melts through the
conventional first-order phase transition. The low density melting
corresponds to the KTHNY scenari \cite{halpnel79}. This kind of
behavior can be understood from the consideration of the potential
(\ref{3}). It is widely believed that the 2D melting transition
scenario corresponds to the KTHNY one for the softer potentials,
however, the systems with hard potentials melt through first-order
transition. The behavior of the system described by the potential
(\ref{3}) is determined by the soft long-range part of the
potential at low densities. At the same time, the hard core of the
potential plays the main role at the high densities. It seems that
this is the reason of the observed peculiarities of the phase
diagram. It was also shown, that the order of the region of
anomalous diffusion and the regions of density and structural
anomalies are inverted in comparison with the $3D$ case and have
silicalike sequence. These results may be also useful for the
qualitative understanding the behavior of confined monolayers of
water confined between two hydrophobic plates
\cite{rev1,rice,barbosa,buld2d,krott}.

\bigskip

We are grateful to S. M. Stishov, V. V. Brazhkin, and E.E.
Tareyeva for stimulating discussions. Yu.F. and E.T. also thank
the Russian Scientific Center Kurchatov Institute and Joint
Supercomputing Center of the Russian Academy of Science for
computational facilities. The work was supported in part by the
Russian Foundation for Basic Research (Grants No 11-02-00341,
13-02-12008, 13-02-00579, and 13-02-00913)  and the Ministry of
Education and Science of Russian Federation (project
MK-2099.2013.2).

\bigskip

\end{document}